\documentclass[twocolumn,showpacs,amsmath,amssymb,prc,floatfix]{revtex4}
\usepackage{graphicx}
\usepackage{dcolumn}
\usepackage{bm}

\begin{document}

\title{Asymptotic High Energy Total Cross Sections\\ 
and Theories with Extra Dimensions}

\author{J. Swain and A. Widom}
\affiliation{Physics Department, Northeastern University, Boston MA USA}
\author{Y.N. Srivastava}
\affiliation{Physics Department \& INFN, University of Perugia, Perugia IT}

\begin{abstract}
The rate at which cross sections grow with energy is sensitive to the presence of
extra dimensions in a rather model-independent fashion. We examine how rates
would be expected to grow if there are more spatial dimensions than 3 which 
appear at some energy scale, making connections with black hole physics
and string theory. We also review what is known about the corresponding
generalization of the Froissart-Martin bound and the experimental status of high energy
hadronic cross sections which appear to saturate it up to the experimentally
accessible limit of 100 TeV. We discuss how extra dimensions can be searched
for in high energy cross section data and find no room for large extra dimensions
in present data. Any apparent signatures of extra dimensions at the
LHC would seem to have to be interpreted as due to some other form of new physics.
\end{abstract}


\maketitle

\section{Introduction \label{intro}}
There is a wealth of possible constraints on new physics which
can be obtained without the need to look for specific exclusive
processes which may be difficult to detect and distinguish from
backgrounds simply by looking at how cross sections grow
with energy. We review simple arguments for how cross sections
are expected to grow with energy, and connect them with black
hole thermodynamics and string-theoretical models. Allowing
for a number of space dimensions greater than 3 we find, in 
agreement with earlier discussions of the Froissart-Martin bound in
higher dimensions, much faster growth of cross sections with energy than
is allowed by unitarity in 3+1 dimensions. Since the experimental results saturate
the unitarity bounds, we find there is no room for extra dimensions\cite{extradims}
at scales below 100 TeV and point out that the best way to 
search for such dimensions may well be via measurements of
the energy dependence of cross sections rather than via searches
for exclusive processes.

\section{High Energy Total Cross Section \label{HETCS}}

A clear physical argument\cite{Eden:1967} for predicting the high center of mass 
energy \begin{math} E \end{math} total cross 
section is the following: (i) If the elastic scattering amplitude at high energy is 
dominated by the exchange of the lightest mass \begin{math} \mu \end{math} particle, 
then the probability of the exchange in space reads 
\begin{equation}
P(r,E)\sim \exp\left[-\frac{2\mu cr}{\hbar}+\frac{S(E)}{k_B}\right],
\label{HETCS1}
\end{equation}
where \begin{math} 2\mu r/\hbar  \end{math} is the WKB factor for a particle to 
move a space-like distance \begin{math} r \end{math} and the {\em entropy} 
\begin{math} S(E) \end{math} determines the density of final states. 
(ii) The probability becomes of order unity at a distance 
\begin{equation}
R=\frac{\hbar S(E)}{2\mu c k_B}\ .
\label{HETCS2}
\end{equation} 
(iii) The total cross section for {\em shadow scattering} is then given by 
\begin{math} \sigma_{\rm total}=2\pi R^2  \end{math} yielding
\begin{equation}
\sigma_{\rm total}(E)=\frac{\pi}{2}\left[\frac{\hbar}{\mu c}\right]^2
\left[\frac{S(E)}{k_B}\right]^2.
\label{HETCS3}
\end{equation} 
The asymptotic \begin{math} E\to \infty  \end{math} total cross section is thereby 
determined by the entropy \begin{math} S(E) \end{math}.

A typical entropy estimate may be made via the following reasoning: The equipartition 
theorem for a gas of ultra-relativistic particles implies a mean particle energy varying 
linearly with temperature;
\begin{math} \overline{\epsilon}\approx \overline{c|{\bf p}|} \approx 3k_BT  \end{math}.  
A Boltzmann gas of such particles has a constant heat capacity.
A system with a constant heat capacity \begin{math} C_\infty \end{math} obeys  
\begin{equation}
E=C_\infty T=C_\infty \frac{dE}{dS}\ \ \ \Rightarrow 
\ \ \ dS=C_\infty\frac{dE}{E}\ ,
\label{HETCS4}
\end{equation} 
yielding an entropy logarithm 
\begin{equation}
S(E)=C_\infty \ln\left(\frac{E}{E_0}\right).
\label{HETCS5}
\end{equation} 
By virtue of Eqs.(\ref{HETCS3}) and (\ref{HETCS5}),  
the total cross section for a constant heat capacity system, 
\begin{equation}
\sigma_{\rm total}(E)=
\frac{\pi}{2}\left[\frac{\hbar C_\infty}{\mu ck_B}\right]^2
\ln^2\left(\frac{E}{E_0}\right),
\label{HETCS6}
\end{equation} 
saturates the Froissart-Martin bound\cite{Froissart:1961,Martin:1963}. 
In a more general thermodynamically stable situation, the entropy 
\begin{math} S(E) \end{math} is determined parametrically by the heat 
capacity as a function of temperature 
\begin{eqnarray}
C(T)=\frac{dE(T)}{dT}=T\frac{dS(T)}{dT} \ ,
\nonumber \\ 
E=\int_0^T C(T^\prime )dT^\prime ,
\nonumber \\ 
S=\int_0^T C(T^\prime )\frac{dT^\prime}{T^\prime}\ . 
\label{HETCS7}
\end{eqnarray} 
The saturation Eqs.(\ref{HETCS5}) and (\ref{HETCS6}) will then hold true 
only in the high anergy and high temperature limit of a stable heat capacity 
\begin{math} C(T\to \infty)=C_\infty  \end{math} .

To compute the total high energy cross section for models with extra 
dimensions, the central theoretical problem is to understand the entropy 
implicit in such models. In so far as extra dimensional theories are 
thought to describe the gravitational interaction, the resulting entropy 
may be expected to exhibit a {\em second law violation}. Let us first review this 
second law instability for the well known case of the black hole entropy.

\section{Black Hole Entropy \label{BHE}}

The gravitational coupling strength of a black hole with a mass 
\begin{math} M \end{math} may be defined via 
\begin{equation}
\alpha_G(M^2)=\frac{GM^2}{\hbar c}=\frac{S}{4\pi k_B}\ ,
\label{BHE1}
\end{equation}
wherein \begin{math} S \end{math} is the black hole entropy, 
\begin{math} E=Mc^2  \end{math} is the black hole energy and  
\begin{equation}
S=4\pi k_B \left(\frac{G}{\hbar c^5}\right)E^2 .
\label{BHE2}
\end{equation}
The black hole temperature is then 
\begin{equation}
\frac{1}{T}=\frac{dS}{dE}
\ \ \ \Rightarrow\ \ \ E=2\pi \tau_G\left(\frac{\hbar c}{k_BT}\right),
\label{BHE3}
\end{equation}
where the gravitational vacuum tension \begin{math} \tau_G  \end{math} 
is determined by 
\begin{equation}
2\pi \tau_G=\frac{c^4}{8\pi G}\approx 4.816\times 10^{42}\ {\rm Newton}.
\label{BHE4}
\end{equation}
The physical meaning of the vacuum gravitational tension  
becomes evident if the Einstein field equations for the energy-pressure 
tensor are written in the form 
\begin{equation}
T_{\mu \nu }=2\pi \tau_G\left[R_{\mu \nu}-\frac{1}{2} g_{\mu \nu }R\right].
\label{BHE5}
\end{equation}
The Planck mass \begin{math} M_P  \end{math} is defined via Eq.(\ref{BHE1}) 
employing \begin{math} \alpha_G(M_P^2)=1 \end{math}, 
\begin{equation}
M_P=\sqrt{\frac{\hbar c}{G}}\approx 6.88\times 10^{-5}{\rm gm}
\approx 3.86\times 10^{28}{eV/c^2}.
\label{BHE6}
\end{equation}
Note that the energy, 
\begin{equation}
\frac{E}{M_Pc^2}=\frac{M_Pc^2}{8\pi k_BT},
\label{BHE7}
\end{equation}
decreases as the temperature increases in violation of the second law of 
thermodynamics, i.e. the black hole heat capacity 
\begin{math} C=dE/dT<0 \end{math}. This instability is a common feature of 
classical self-gravitationally bound systems.

\section{Hagedorn-String Entropy \label{TED}}

Theories with extra dimensions grew out of string theories of interactions 
which would include in principle gravitational theories. A classical picture 
of a rotating string can be developed. If one considers 
a string of length \begin{math}L \end{math}, then the lowest resonance
frequency obeys 
\begin{equation}
\omega = \frac{\pi c}{L}\ .
\label{TED1}
\end{equation} 
Treating the resonant frequency \begin{math} \omega \end{math} 
as an angular velocity conjugate to the string angular momentum 
\begin{math} J \end{math} one finds 
\begin{equation}
\omega = \frac{dE}{dJ}
\label{TED2}
\end{equation} 
where the energy is related to the string tension 
\begin{math} \tau_S  \end{math} via 
\begin{equation}
E=\tau_S L.
\label{TED3}
\end{equation} 
Eqs.(\ref{TED1}), (\ref{TED2}) and (\ref{TED3}) imply the differential 
equation, 
\begin{equation}
dJ=\frac{EdE}{\pi c \tau_S}\ ,
\label{TED4}
\end{equation} 
whose solution is the linear Regge trajectory, 
\begin{equation}
J=\hbar \left(\alpha_0+\alpha^\prime E^2\right).
\label{TED5}
\end{equation} 
In Eq.(\ref{TED5}), \begin{math} \alpha_0  \end{math} is the 
Pomeranchuck intercept\cite{Pomeranchuck:1956,Pomeranchuck:1958} and 
\begin{equation}
2\pi \tau_S=\frac{1}{\hbar c \alpha^\prime} 
\label{TED6}
\end{equation} 
is the generalized string tension analog of the gravitational tension 
Eq.(\ref{BHE4}) which is now determined by the Regge trajectory slope 
parameter \begin{math} \alpha^\prime  \end{math}. Let us consider the 
entropy consequences of these types of models.

\subsection{Hagedorn Temperature \label{HT}}

The canonical partition function for a system in an environmental 
temperature \begin{math} \tilde{T}  \end{math} has the form 
\begin{equation}
Z(\tilde{T})=\sum_E \Omega (E) \exp(-E/k_B\tilde{T}),
\label{HT1}
\end{equation}
where \begin{math} \Omega (E)  \end{math} is the number of quantum states 
with energy \begin{math} E \end{math}. In terms of the entropy 
\begin{equation}
S(E)=k_B \ln \Omega(E)
\label{HT2}
\end{equation}
and the canonical free energy 
\begin{equation}
F(\tilde{T})=-k_B\tilde{T} \ln Z(\tilde{T}),
\label{HT3}
\end{equation}
one finds from Eqs.(\ref{HT1}), (\ref{HT2}) and (\ref{HT3}) that
\begin{equation}
\exp \left[-\frac{F(\tilde{T})}{k_B\tilde{T}}\right]
=\sum_E \exp 
\left[
\frac{S(E)}{k_B}-\frac{E}{k_B\tilde{T}}.
\right].
\label{HT4}
\end{equation}
The Hagedorn string entropy has the form\cite{Hagedorn:1965,Hagedorn:1968} 
\begin{equation}
S(E)=\frac{E}{T_H}-k_B \eta \ln\left(\frac{E}{k_B T_H}\right)+k_B\gamma ,
\label{HT5}
\end{equation}
wherein \begin{math} \eta \end{math} and \begin{math} \gamma \end{math} 
are dimensionless constants and \begin{math} T_H \end{math} is the 
Hagedorn temperature.

\subsection{Statistical Thermodynamics \label{ST}}

In order that the partition function in Eq. (\ref{HT1}) converge, the environmental 
temperature \begin{math} \tilde{T} \end{math} must be less than the Hagedorn 
temperature \begin{math} T_H \end{math}. 
\begin{equation}
\tilde{T}<T_H\ ,
\label{HT6}
\end{equation}
On the other hand, the string 
temperature \begin{math} T \end{math} given by,
\begin{equation}
\frac{1}{T}=\frac{dS}{dE},
\label{HT7}
\end{equation}
yields a positive energy 
\begin{equation}
E=\eta k_BT_H \left[\frac{T}{T-T_H}\right]
\equiv \Phi \left[\frac{T}{T-T_H}\right] > 0, 
\label{HT8}
\end{equation}
Eq.(\ref{HT8}) implies a string temperature larger than the Hagedorn temperature,
\begin{equation}
T>T_H\ ,
\label{HT9}
\end{equation}
as well as a threshold for the string excitation energy 
\begin{equation}
E>\Phi =\eta k_B T_H,
\label{HT10}
\end{equation}
Note the inequalities 
\begin{equation}
\tilde{T} < T_H\ < T ,
\label{HT11}
\end{equation}
which imply an environmental temperature \begin{math} \tilde{T} \end{math} which is 
less tha the string temperature \begin{math} T \end{math}. 
That a thermodynamically stable environment cannot come 
into equilibrium with a string is due to the latter having negative heat capacity 
\begin{equation}
C=\frac{dE}{dT}=-\Phi \left[ \frac{T_H}{(T-T_H)^2}  \right] <0
\label{HT12}
\end{equation}
which is a thermodynamic second law violation.

\subsection{String Theory\label{BS}}

The extra dimension value \begin{math} n \end{math} for space-time is 
related to the full dimension of space-time \begin{math} \cal{D} \end{math} 
via 
\begin{equation}
{\cal D}=3+1+n=4+n.
\label{BS1}
\end{equation}
The Hagedorn entropy for string theories is illustrated by 
bosonic string example where the entropy can be computed from the number 
theory integer partition functions\cite{Hardy:1960}. It is 
\begin{eqnarray}
\frac{S(E)}{k_B}=2\pi E\sqrt{\alpha^\prime \left[\frac{{\cal D}-2}{6}\right]}
\nonumber \\ 
-\left(\frac{{\cal D}+1}{2}\right)\ln (\sqrt{\alpha^\prime}E)
\nonumber \\
\left(\frac{{\cal D}-1}{4}\right)\ln \left(\frac{{\cal D}-2}{24}\right)
-\frac{\ln 2}{2}\ ,
\label{BS2}
\end{eqnarray}
which is clearly a special case of the general Hagedorn string entropy Eq.(\ref{HT5}).
For the supersymmetric string which includes fermions, the functional form of the 
Hagedorn entropy in Eq.(\ref{HT5}) is expected to remain intact. However, the detailed 
dependence of the parameters \begin{math} T_H, \eta \end{math} and 
\begin{math} \gamma \end{math} on \begin{math} {\cal D} \end{math} is expected to 
depend on the specific string model chosen.

\section{Compact Extra Dimensions\label{CED}}

\begin{figure}[htbp] 
   \centering
   \includegraphics[width=3.5in]{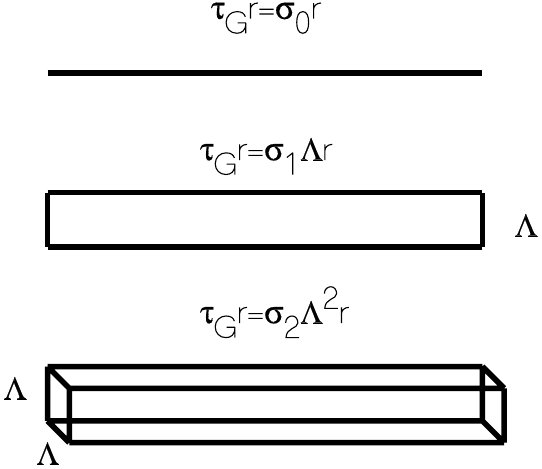} 
   \caption{For a one dimenisional string of length $r$ under gravitational tension, 
$\tau_G \equiv \sigma_0$, the energy of stretching $\tau_G r$ enters into Newton's 
gravitational law as in the denominator of Eq.(\ref{CED1}). If the string has one extra 
dimension of length $\Lambda \ll r$, then the stretching two dimensional membrane is 
described by a surface tension $\sigma_1=\tau_G/\Lambda $. If the string has two extra 
dimensions each of length $\Lambda \ll r$, then the stretching three dimensional square 
prism is described by a negative pressure $\sigma_2=\tau_G/\Lambda^2 $. For the case of 
$n \ge 1$ added dimensions, one finds $\sigma_n=\tau_G/\Lambda^n $ as in Eq.(\ref{CED1}).}
   \label{fig1}
\end{figure}

Let us suppose that \begin{math} n \end{math} extra dimensions are compact 
with a gravitational length scale \begin{math} \Lambda \end{math}. For distance  
\begin{math} r \end{math} large on the length scale \begin{math} \Lambda \end{math}, 
the Newtonian attraction is determined by the gravitational string tension in 
Eq.(\ref{BHE4}); 
\begin{eqnarray}
-U=G\frac{m_1m_2}{r}
\nonumber \\ 
=\frac{1}{16\pi^2}\left[\frac{(m_1c^2)(m_2c^2)}{\tau_G r}\right] 
\nonumber \\
=\frac{1}{16\pi^2}\left[\frac{(m_1c^2)(m_2c^2)}{\sigma_n \Lambda^n r}\right]
\ \ (r\gg \Lambda ).
\label{CED1}
\end{eqnarray}
The nature of adding \begin{math} n \end{math} compact extra dimensions is illustrated 
schematically in FIG.\ref{fig1}. For short distances, 
the power law energy has an exponent 
\begin{equation}
-U\propto \left[\frac{(m_1c^2)(m_2c^2)}{\sigma_n r^{1+n}}\right]\ \ \ (r\ll \Lambda ).
\label{CED2}
\end{equation}

Similar considerations apply for the string entropy with \begin{math} n  \end{math}
extra dimansions each of length \begin{math} \Lambda  \end{math}. For a gravitational 
string one may begin with an entropy of the Hagedorn form as given in 
Eq.(\ref{HT5}); i.e. as \begin{math} E\to \infty  \end{math}
\begin{equation}
\frac{S_0(E)}{k_B}=\varpi_0 \left[\frac{E}{\sqrt{\hbar c\tau_G}}\right]+\cdots ,
\label{CED3}
\end{equation}
wherein \begin{math} \varpi_0 \end{math} is a dimensionless constant of order unity. 
With \begin{math} n \end{math} extra compact dimensions 
\begin{eqnarray}
\frac{S_n(E)}{k_B}=\varpi_n \left[\frac{E}{\sqrt{\hbar c\sigma_n}}\right]
\left(\frac{E}{\hbar c}\right)^{n/2}+\cdots 
\nonumber \\ 
=\varpi_n \sqrt{\frac{\hbar c}{\sigma_n}}
\left(\frac{E}{\hbar c}\right)^{1+(n/2)}+\cdots . 
\label{CED4}
\end{eqnarray}

This can be compared with the string scattering results of Amati, Ciafaloni and
Veneziano who find for the total, diffractive and inclusive cross sections $\sigma_{tot}$,
$\sigma_{dif}$ and $\sigma_{inel}$ respectively

\begin{math}
\sigma_{tot} \sim s^{\frac{D-2}{D-4}}\\
\sigma_{dif} \sim s\\
\sigma_{inel} \sim (\ln(s))^{D-2}\\
\end{math}

\noindent at high energies, where $D$ is the number of dimensions of spacetime.

We now discuss how extra compact dimensions have an effect on the total hadron-hadron 
cross section in the linit of high center of mass energy.

\section{Total Hadronic Cross Section \label{TCS}}

From the relationship between the total cross section entropy in Eqs.(\ref{HETCS3}), 
(\ref{HETCS5}) and (\ref{CED4}), one finds the total cross section in the 
high energy limit 
\begin{eqnarray}
\sigma_{\rm total}(s)=\frac{\pi}{2}\left[\frac{\hbar}{\mu c}\right]^2 \times
\nonumber \\ 
\left|\frac{C}{k_B}\ln\left(\frac{s}{s_0}\right)+ 
\varpi_n \sqrt{\frac{\hbar c}{\sigma_n}}\ s^{(n+2)/4}+\cdots  \right|^2
\label{TCS1}
\end{eqnarray}
wherein the center of mass energy 
\begin{math} E=\hbar c \sqrt{s}\to \infty \end{math}.
Below the threshold for the observation of extra dimensions  
\begin{eqnarray}
\left[s_0\ll s \ll \left(\frac{\sigma_n}{\hbar c}\right)^{2/(2+n)}\right],
\nonumber \\ 
\sigma_{\rm total}(s)=\frac{\pi}{2}\left[\frac{C\hbar}{k_B\mu c}\right]^2
\ln^2\left(\frac{s}{s_0}\right).
\label{TCS2}
\end{eqnarray}
Above the threshold for the observation of extra dimensions,
\begin{eqnarray}
\left[\left(\frac{\sigma_n}{\hbar c}\right)^{2/(2+n)}\ll s\right],
\nonumber \\ 
\sigma_{\rm total}(s)=\frac{\pi \varpi_n^2\hbar c}{2\sigma_n}
\left[\frac{\hbar}{\mu c}\right]^2 s^{(n+2)/2}.
\label{TCS3}
\end{eqnarray}
The transition from Eq.(\ref{TCS2}) to (\ref{TCS3}) at the energy 
\begin{eqnarray}
E_{\rm transition} \sim \hbar c\left(\frac{\sigma_n}{\hbar c}\right)^{1/(2+n)}
\nonumber \\ 
\sim \left(\frac{\tau_G }{\hbar c}\right)^{1/(2+n)}\frac{\hbar c}{[\Lambda^n]^{1/(n+2)}}
\label{TCS4}
\end{eqnarray}
which depends on both the number \begin{math} n \end{math} of extra dimensions and the 
compact length scale \begin{math} \Lambda \end{math}.

\section{The Froissart Bound in Higher Dimensions}

The discussions above have been quite physical, but one can also directly generalize
the Froissart bound by repeating the original argument but now in more than 3 spatial 
dimensions. This has been done by Chaichian and Fisher\cite{Chaichian:1988} who
find that cross sections must be bounded by higher powers of $\ln(s)$ or even
as powers of $s$ multiplied by logarithms.

\section{Generic Effects of Higher Dimensions}

A key observation of this paper is that if one assumes that there are extra dimensions of
space, one generically expects a qualitative change in how cross sections rise with 
energy. That change is that they rise significantly faster than one would expect in
3 spatial dimensions as soon energies rise high enough that those extra dimensions
are accessible. This is an essentially model-independent statement, and due physically
to the fact the adding dimensions opens up phase space for decays. This is true
whether or not the Standard Model matter particles are restricted to lie on some
3-dimensional ``brane''\cite{extradims}. As long as one can excite degrees of freedom which
see more than 3 space dimensions, the rate at which cross sections grow will be
higher than one would expect in 3. Note that such behavior could even 
be expected simply on dimensional grounds, with the material presented above
being a clarification of this point.

Now it is well-known that hadronic cross sections saturate the unitarity bounds so one
could search for evidence of extra dimensions simply by search for energies above
which cross sections grow faster than with energy than at lower dimensions.  A detailed
fit with two functional forms above and below $E_{transition}$ as defined above could
be used to clarify the nature and number of such extra dimensions, should any evidence
for them at all arise. However, as we show
in the following section, extra dimensions seem to be ruled out up to scales of about 100 TeV 
already from the present data.

\section{Constraints on compact dimensions from high energy cross-section data \label{Cons}}

The experimental situation on high-energy hadronic cross-sections may be summarized as follows. High quality data exist up to $\sqrt{s}\ =\ 2\ $TeV for $\sigma_{total}^{(p\bar{p})}(s)$  from the Tevatron and up to $\sqrt{s}\ =\ 0.2\ $TeV for $\sigma_{total}^{(pp)}(s)$. Less precise data from cosmic rays are available for $\sigma_{total}^{(pp)}(s)$ up to $\sqrt{s}\ =\ 100\ $TeV. Recently, the ATLAS group at LHC has released preliminary data on $\sigma_{inelastic}^{(pp)}(s)$ at $\sqrt{s}\ =\ 7\ $TeV.

There are three reasonably complete phenomenological analyses of the above cross-section data which may be summarized as follows:

\begin{itemize}
\item PDG\cite{PDG}:  They {\it assume} that $\sigma_{total}(s)$ rises as $(ln(s/s_o))^2$, the {\it maximum} rise allowed by the Froissart bound. Their parametrization for all the available data reads
\begin{eqnarray}
\sigma_{total}^{(pp,p\bar p)}(s) = (0.308 {\rm mb})\ [\ln(\frac{s}{{\rm GeV}^2})]^2\nonumber\\ 
+ (35.45 {\rm mb}) +..., 
\label{PDG}
\end{eqnarray}
the dots meaning non-leading terms vanishing at high energy. A $\chi^2/{\rm d.o.f.}\ =\ 1.05$ -for all data considered- is given.

\item BH\cite{Block}: Block and Halzen also find  strong evidence that the Froissart bound is saturated. That is, their result for a generic hadronic cross-section (in the asymptotic domain) may be written as
\begin{eqnarray}
\sigma_{total}^{(pp,p\bar p)}(s) = C_1 [\ln(\frac{s}{{\rm GeV}^2})]^2\nonumber\\ 
+\ C_2 [\ln(\frac{s}{{\rm GeV}^2})]\nonumber\\ 
+ C_3 +..., 
\label{BH}
\end{eqnarray} 
where the constants $C_i,\ i=1,2,3$ are determined from the data.

\item GGPS\cite{GGPS}: This is an independent theoretical analysis -developed over two decades- which is based on mini-jets, incorporating aspects of confinement and soft-gluon resummation. It has been quite successful in obtaining a rather complete description of total cross-sections for $pp, p\bar{p}, \pi p, \gamma p, \gamma\gamma$ processes. This group finds the rise to be $(\ln(s/s_o))^{q}$ where $1<q<2$, with a phenomenological value: $4/3<q<3/2$. 
\end{itemize}

Thus, {\it all} viable phenomenological analyses of total and inelastic cross-sections  up to $\sqrt{s}\ =\ 100\ $TeV conclude that
\begin{equation}
\label{P1}
\sigma(s)\ \rightarrow\ \sigma_o [ln(s/s_o)]^q, \ \ \ with\ q\leq\ 2.
\end{equation}      
Eq.(\ref{P1}) is at complete odds with Eq.(\ref{TCS3}) even for $n\ =\ 1$. Hence, there appears to be no room for any compactified dimensions whose thresholds are below $100\ $TeV.

\begin{figure}
\scalebox {0.5}{\includegraphics{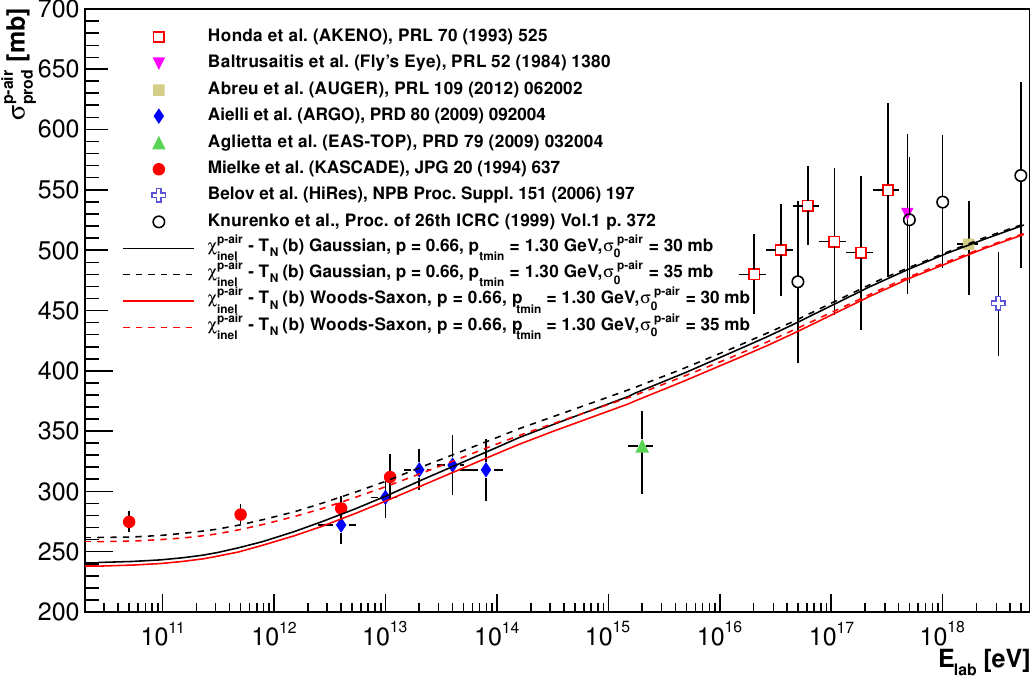}}
\caption{A plot of high energy cosmic ray data from\cite{Honda, Baltrusaitis, Abreu, Aielli, Aglietta, 
Mielke, Belov, Knurenko} for $p-air$ cross-section are presented along
with a theoretical model for the same obtained using input that fits the $pp$ cross-section\cite{Grau}} ({\em q.v.}).

\label{xsec}
\end{figure}

The good agreement between the observed behavior of high energy cosmic ray cross section data and that of
cross section data obtained from $pp$ collisions at lower energies obtained in laboratory experiments
can be seen in graphically in Figure \ref{xsec}.
The highest $pp$ scattering data from LHC extends up to $\sqrt{s} = 8\ TeV$ and has not found evidence
for large extra dimensions \cite{CMSrecent,ATLASrecent}.

All theoretical analyses
confirm that the Froissart bound (as obtained in $D = 4$) is obeyed with no need of any contributions from
higher dimensions. Furthermore, the most recent accurate cosmic ray $p-air $ data from the Auger Collaboration
\cite{Abreu} extending up to $\sqrt{s} = 57\ TeV$ have also been analyzed using the same model employed to 
describe the LHC $pp$ data and once again there is no evidence of higher dimensions. See Figure \ref{xsec}.
Note that it is not a fit to the AUGER data but simply an extrapolation to higher energies from the LHC data.

\section{Conclusions}

We argue on general grounds, and via concrete calculations, that if extra space dimensions are
accessible to at least some excitations produced in high energy hadronic collisions above some energy
scale $E_{critical}$ one should see an increase in the rate at which total cross sections rise. This
suggests a general means to search for such extra dimensions by looking at cross sections
as a function of energy. Such searches are particularly attractive since they are highly model-independent
and largely unaffected by single events which might, due to statistical fluctuations, seem to support
some model with higher dimensions. 
Since the data so far apparently saturate the Froissart-Martin bound in 3+1 dimensions,
there appears to be no room for any compactified dimensions whose thresholds are below $100$ TeV. 
This limit can be extended as data from ever higher energy collisions are obtained, but for the near
future this will have to come from cosmic ray data. Any events which might seem to be signals of
extra spacetime dimensions at the LHC would have to be attributed to some other sort of new physics.

Since this paper was written, Block and Halzen \cite{BH1}(who have accepted our work with an aim to
extending it) have suggested that 
higher dimensions might be ruled out to arbitrarily high energies via the same argument, but they use
a Froissart bound which is tied to 3+1 dimensions, which as discussed above, is dimension-dependent. Further recent discussions 
include that of Fagundes, Menon and Silva \cite{FMS1,FMS2} (see also the commentary by Block and Halzen\cite{BH2})
who suggest that energy dependence of hadronic total cross section at high energies may be an open problem
still). Block and Halzen have also recently argued that the Auger cross section data supports the idea that
proton asymptotically develops into a disk\cite{BH3}.

None of these publications changes the conclusions of our paper.

\section{Acknowledgements}

J. S. thanks National Science Foundation for its support via NSF grants PHY-0855388 and
PHY-1205845.

\end{document}